\documentclass{iaus}
\usepackage{graphicx}

\title[Galactic disk modelling] 
{The Challenge of Modelling Galactic Disks}

\author[Andreas Burkert]   
{Andreas Burkert$^1$}

\affiliation{$^1$University Observatory, University of Munich, \\ Scheinerstr. 1, D-81679, Munich, Germany
\\ email: {\tt burkert@usm.lmu.de}}

\pubyear{2008}
\volume{254}  
\pagerange{119--126}
\jname{The Galaxy Disk in Cosmological Context}
\editors{J. Andersen \& B. Nordstr\"om, eds.}
\begin{document}

\maketitle

\begin{abstract}
Detailed models of galactic disk formation and evolution require
knowledge about the initial conditions under which disk galaxies form,
the boundary conditions that affect their secular evolution and
the micro-physical processes that drive the multi-phase interstellar medium
and regulate the star formation history.
Unfortunately, up to now, most of these ingredients are
still poorly understood. The challenge therefore is to, despite this caveat,
construct realistic models of galactic disks with predictive
power. This short review will summarize some problems related to
numerical simulations of galactic disk formation and evolution.

\keywords{Galaxy: disk, galaxies: formation, galaxies: evolution, 
ISM: general, ISM: clouds, hydrodynamics, turbulence}
\end{abstract}

\firstsection 
\section{Initial- and Boundary Conditions: The Cosmological Angular Momentum Problem}

The radial surface density profiles of galactic disks are determined
by the gravitational potential of the galaxy that is dominated in the outer parts
by dark matter and the specific angular momentum
distribution of the infalling gas that dissipates its potential and kinetic
energy while settling into centrifugal equilibrium in the inner regions
of a dark matter halo. In addition one has to consider the secular evolution 
of galactic disks. Viscous angular momentum redistribution and
selective gas loss in galactic winds strongly affects the evolution of disks 
making it difficult to infer the initial conditions from their presently observed  structure.

Angular momentum is an important ingredient in order for
galactic disks to form. It is generally assumed that, before collapse,
gas and dark matter are well mixed
and therefore acquire a similar specific angular momentum distribution
(Peebles 1969; Fall \& Efstathiou 1980; White 1984). If angular momentum would be conserved during gas
infall, the resulting disk size should be directly related to the 
specific angular momentum $\lambda'$ of the surrounding dark halo where
(Bullock et al. 2001)

\begin{equation}
\lambda' = \frac{J}{\sqrt{2} M_{vir} V_{vir} R_{vir}}
\end{equation}

\noindent with $R_{vir}$ and $V_{vir}^2=GM_{vir}/R_{vir}$ the virial radius and 
virial velocity of the halo, respectively, and $M_{vir}$ its virial mass.
Adopting a flat rotation curve, the disk scale length is
(Mo et al. 1998; Burkert \& D'Onghia 04)

\begin{equation}
R_d \approx 8 \left( \frac{\lambda'}{0.035} \right) \left( \frac{v_{max}}{200 km/s} \right) kpc
\end{equation}

\noindent where $v_{max}$ is the maximum rotational velocity in the disk. 

\begin{figure}[t]
\begin{center}
 \includegraphics[width=3.4in]{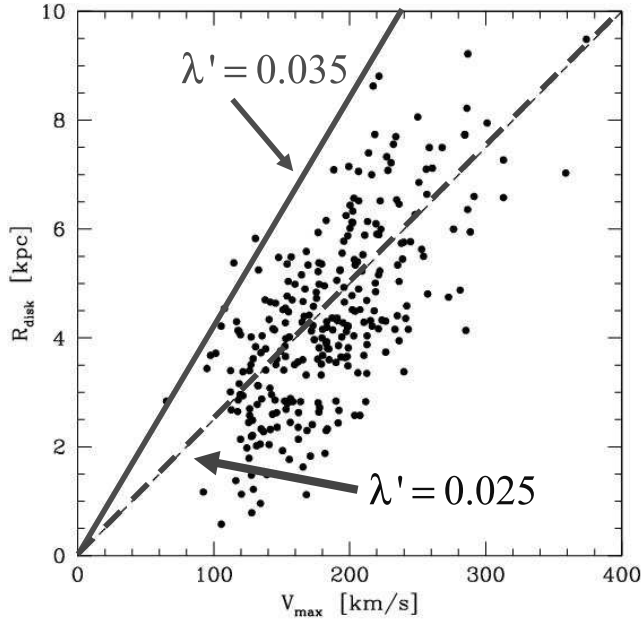} 
 \caption{The observed scale lengths versus the maximum rotational velocities of galactic disks are
shown for the Courteau (1997) sample. The sold line shows the theoretically predicted correlation
for $\lambda' = 0.035$. The dashed curve corresponds to $\lambda'=0.025$.}
   \label{fig1}
\end{center}
\end{figure}

Figure 1 shows the correlation between the disk scale length
$R_{disk}$ and the maximum rotational velocity $v_{max}$ for massive spiral galaxies (Courteau 1997) which is 
consistent with a mean value of $\lambda' \approx 0.025$. 
The observationally derived average peak value
of $\lambda' = 0.025$ is somewhat smaller than the theoretically predicted value
of $\lambda' = 0.035$,  indicating that the gas 
could on average have lost some amount of angular momentum during the phase of
decoupling from the dark component and settling into the equatorial plane.

This result is promising. The situation is however far less satisfactory when we 
consider more detailed numerical models of gas infall and accumulation in galactic disks.
Many simulations of galactic disk formation suffer from 
catastrophic angular momentum loss which 
leads to disks with unreasonably small scale lengths and surface densities that are too large.
The origin of this problem has been attributed to a strong clumping of the infalling gas 
which looses angular momentum by dynamical friction within the surrounding dark matter halo 
(Navarro \& Benz 1991, Navarro \& Steinmetz 2000). Other possibilites are low numerical
resolution (Governato et al. 2004, 2007), the effect of substantial and major mergers (d'Onghia et al. 2006) 
or artificial secular angular momentum transfer from the cold disk to its hot surrounding 
(Okamoto et al. 2003). Over the years many groups have tried to solve this problem by
including star formation and energetic feedback (e.g. Sommer-Larsen et al. 2003, Abadi et al. 2003,
Springel \& Hernquist 2003, Robertson et al. 2004, Oppenheimer \& Dave 2006, Dubois \& Teyssier 2008).
The results are however confusing. First of all the origin of the angular momentum problem
is not clearly understood. Secondly, no reasonable, universally applicable feedback 
prescription has been found that would lead to the formation of large-sized, 
late-type disks, not only for one special case, but in general.

Progress in our understanding of the cosmological angular momentum problem
has recently been achieved by Zavala et al. (2008) who confirmed
that the specific angular momentum distribution
of the disk forming material follows closely the angular momentum evolution of the dark matter halo.
The dark matter angular momentum grows at early times as a result of large-scale tidal torques, 
consistent with the prediction of linear theory and remains constant after the epoch of maximum 
expansion. During this late phase however angular momentum is redistributed
within the dark halo with
the inner dark halo regions loosing up to 90\% of their specific angular momentum to
the outer parts.
The process leading to this angular momentum redistribution is not
discussed in details. It is however likely that substantial 
mergers with mass ratios less than 10:1 that are expected to occur frequently even at late
phases during galaxy formation perturb the halo and 
generate global asymmetries in the mass distribution that are known to
be an efficient mechanism for angular momentum transfer. Small satellite infall is probably of
minor importance. It would be interesting to study the role of major and minor mergers
in this process in greater details.

It is then likely that any gas residing in the inner regions during such
an angular momentum redistribution
will also loose most of its angular momentum, independent of whether the gas resides already
in a protodisk, is still confined to dark matter substructures or is in an extended,
diffuse distribution.
Zavala et al. (2008) (see also Okamoto et al., 2005 and Scannapieco et al., 2008) show that
efficient heating of the gas component can prevent angular momentum loss, probably
because most of the gaseous component resides in the outer parts of the dark halo
during its angular momentum redistribution phase. The gas would then actually gain angular
momentum  rather than loose it and could lateron settle smoothly into
an extended galactic disk in an ELS-like (Eggen, Lynden-Bell \& Sandage 1962) accretion phase.

Unfortunately little is known about the energetic processes that
could lead to such an evolution. Obviously, star formation must be delayed during the
protogalactic collapse phase in order 
for the gas to have enough time to settle into the plane before condensing into stars. 
However star formation is also required in order to heat the gas preventing
it from collapsing prior to the angular momentum redistribution phase.
Scannapieco et al. (2008)
show that their supernova feedback prescription is able to regulate star formation while at the same time pressurizing
the gas. Their models are however still not efficient enough in order to produce disk-dominated,
late-type galaxies. Large galactic disks are formed. The systems are however dominated by a
central, massive, low-angular momentum stellar bulge component. This is in contradiction with
observations which indicate a large fraction of massive disk galaxies with bulge-to disk
ratios smaller than 50\% (Weinzirl et al. 2008) that cannot be produced currently by numerical simulations
of cosmological disk formation.

\section{Energetic Feedback and Star Formation}

As discussed in the previous section, star formation and energetic feedback plays a dominant
role in understanding the origin and evolution of galactic disks and in determining the
morphological type of disk galaxies. Scannapieco et al. (2008) for example demonstrate that the
same initial conditions could produce either an elliptical or a disk galaxy, depending
on the adopted efficiency of gas heating during the protogalactic collapse phase.
We do not
yet have a consistent model of the structure and evolution of the multi-phase, turbulent 
interstellar medium and its condensation into stars. This situation is now
improving rapidly due to more sophisticated numerical methods and fast computational
platforms that allow us to run high-resolution models, incorporating a large number of possibly
relevant physical processes (Wada \& Norman 2002, Krumholz \& McKee 2005, Tasker \& Bryan 2008, 
Robertson \& Kravtsov 2008).
Most cosmological simulations however have up to know adopted simplified
observationally motivated descriptions of star formation that are based
on the empirical Kennicutt relations (Kennicutt 1998, 2007) that come in two different version.
The first relation (K1) represents a correlation between the star formation rate per surface area
$\Sigma_{SFR}$ and the gas surface density $\Sigma_g$, averaged over the whole galaxy

\begin{equation}
\Sigma_{SFR}^{(K1)} = 2.5 \times 10^{-4} \left(\frac{\Sigma_g}{M_{\odot}/pc^2}\right)^{1.4} 
\frac{M_{\odot}}{kpc^2 \ yr}
\end{equation}

\noindent The second relation (K2) includes a dependence on the typical orbital period
$\tau_{orb}$ of the disk

\begin{equation}
\Sigma_{SFR}^{(K2)} = 0.017 \left(\frac{\Sigma_g}{M_{\odot}/pc^2}\right)
\left( \frac{10^8 yrs}{\tau_{orb}} \right) \frac{M_{\odot}}{kpc^2 \ yr}.
\end{equation}

\noindent These relationships have been derived from observations as an average over
the whole disk. They are however often also used as theoretical prescriptions for the local
star formation rate which appears observationally justified if 
the total gas surface densitiy $\Sigma_g$ is replaced by the local surface density of molecular gas.
The origin of both relationships is not well understood yet. For example,
Li et al. (2005, 2006) 
ran SPH simulations of a gravitationally unstable gaseous disks, confined by the gravitational
potential of a surrounding dark matter halo. Gravitationally bound gas clumps form in their disks
and are replaced by accreting sink particles. The authors assume that 30\% of the mass of these particles 
is in stars with the rest remaining gaseous. However, no stellar feedback or a destruction mechanism of 
the partly gaseous sink particles was adopted. The star formation surface density is investigated
for different galactic disk models with different rotational velocities and initial gas surface densities.
The authors find a good agreement with the first Kennicutt relation (K1) if they correlate 
$\Sigma_{SFR}$ with $\Sigma_g$ 
at a time when the star formation rate has decreased by a factor of 2.7 with respect to the
initial value which in their model typically corresponds to an evolutionary time of a few $10^7$ yrs.
The significance of this result is however not clear. Obviously, the galaxies studied by Kennicutt are much older
and in a phase of self-regulated star formation that cannot be considered in models without
energetic feedback.  In addition, the authors cannot reproduce the second relation (K2), indicating 
that K2 is not directly related to K1 but instead represents a second constraint for theoretical models.

We can combine K1 and K2 and derive a relationship between the average gas density in 
galactic disks and their orbital period

\begin{equation}
\Sigma_g \sim \tau_{orb}^{-2.5} \sim \left( \frac{v_{rot}}{R_{disk}} \right)^{2.5}
\end{equation}

\noindent where $v_{rot}$ and $R_{disk}$ are the rotational velocity and the size of the galactic disk,
respectively.
This result is puzzling as it is not clear why the kinematical properties of galactic disks
should correlate with their gas surface densities especially in galaxies of Milky Way type or earlier 
where the gas fraction is small compared to the mass in stars. 
Recent detailed hydrodynamical simulations of disk galaxies by Robertson \& Kravtsov (2008), including
low-temperature gas cooling and molecular hydrogen physics can indeed reproduce both Kennicutt relations.
The authors however note themselves that the physical reason for the origin of the K2-relation in their 
simulations is unclear. They argue that in disk galaxies with exponential density profiles
the disk surface density should scale with the orbital period as $\Sigma_d \sim \tau_{orb}^{-2}$.
In this case, K2 requires that $\Sigma_g \sim \Sigma_d^{1.2} \sim (\Sigma_* + \Sigma_g)^{1.2}$
with $\Sigma_*$ the stellar surface density. It is not clear why this relation should hold, especially
for disks with $\Sigma_* > \Sigma_g$.

\section{Secular Evolution and Turbulence in Galactic Disks}

Bullock et al (2001) demonstrated that dark halos have a universal
angular momentum distribution that should also be characteristic for the infalling gas component.
Van den Bosch et al. (2001) lateron showed that this angular momentum distribution is
not consistent with the observed distribution of exponential galactic disks indicating that
viscous angular momentum redistribution in galactic disks must have played an important role.
The viscosity is likely driven by interstellar turbulence which is a result of
stellar energetic feedback processes (see Fig. 2) or global disk instabilities (magneto-rotational instability
or gravitational instability).
Note, that viscous effects will increases the angular momentum problem substantially as viscosity in general
removes angular momentum from the dominante mass component in the disk and transfers it to the outermost
parts of the disk.

\begin{figure}[t]
 \vspace*{1.0 cm} 
\begin{center}
 \includegraphics[width=5.4in]{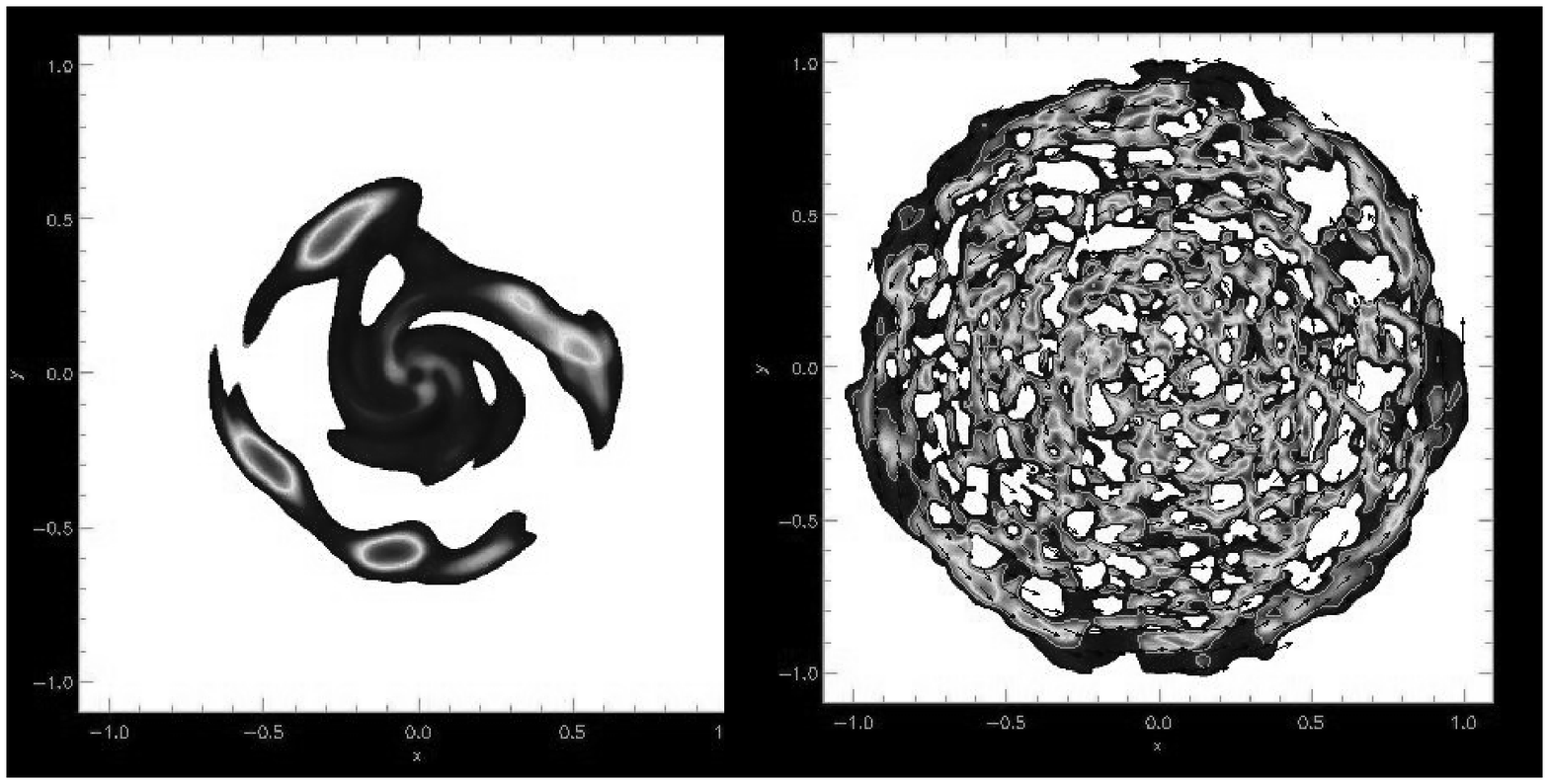} 
 \caption{Gas surface density of a gravitationally unstable gas-rich galactic disk, 
embedded in a dark matter halo.  The left panel shows the gas density distribution 
if star formation is suppressed. The disk forms a few massive gaseous clumps that
spiral into the center by dynamical friction. The situation is however very different
if star formation and stellar energy feedback is included. In this run, supernova explosions
efficiently disrupt dense clumps before they can merge into giant cloud complexes while at
the same time generating a highly turbulent and filamentary multi-phase interstellar medium
(Burkert et al. 2009).}
   \label{fig2}
\end{center}
\end{figure}

The viscous formation of exponential stellar disks from gas disks with various different surface density
distributions has been studied e.g. by Slyz et al. (2002). Their numerical simulations show that
exponential disks form if the star formation timescale is of order
the viscous timescale. Genzel et al. (2008) derive a timescale for turbulent viscosity in galactic
disks of

\begin{equation}
\tau_{visc} = \frac{1}{\alpha} \left( \frac{v_{rot}}{\sigma} \right)^2 \tau_{orb}
\end{equation}

\noindent where $\alpha$ is of order unity. $\tau_{visc} \approx 10^{10}$ yrs for
disks like the Milky Way with $\sigma \approx$ 10-20 km/s and self-regulated low star 
formation rates. H$\alpha$ integral field
spectroscopy has however  detected $z \sim 2$ star forming disk galaxies with large random
gas motions of order 40 km/s to 60 km/s and viscous timescales of less than $10^9$yrs
(Genzel et al., 2006, 2008, F\"orster-Schreiber et al. 2006).
Interestingly, for these objects, the star formation timescales are again similar to the
viscous timescale, leading to star formation rates of 100 M$_\odot$/yr and confirming
that galactic disk gas turbulence, star formation and secular evolution are intimately coupled.
The origin of the clumpiness and high turbulence in redshift 2 disks is not well understood
yet. It seems likely that it is a result of substantial filamentary gas inflow (Dekel et al. 2008),
combined with gravitational instabilites in the disk (Bournaud et al. 2007).

Turbulence seems to regulate star formation not only on large galactic scales but also on local
cloud scales. Most of the molecular gas in the Milky Way is found in giant molecular clouds
with masses of order $10^4-10^6 M_{\odot}$, temperatures of order 10 K and average densities
of order 100 $cm^{-3}$. As their Jeans mass is of order 20 M$_{\odot}$ which is much smaller
than their total mass one would expect that molecular clouds should collapse and condense into stars
on a local free-fall time which is of order $5 \times 10^6$ yrs. Adopting 
a total molecular mass of $M_{H_2} \approx 3 \times 10^9 M_{\odot}$ 
and assuming that a fraction $\eta_{SF} \approx 0.1$ of
the molecular cloud's mass forms stars, the inferred mean star formation rate in the Milky
Way is

\begin{equation}
SFR = \eta_{SF} \frac{M_{H_2}}{\tau_{ff}} \approx 60 M_{\odot}/yr
\end{equation}

\noindent which is an order of magnitude larger than observed.
A possible solution of this problem is turbulence. Molecular clouds are observed
to be driven and shaped by supersonic turbulence that might strongly affect their stability and
star formation rate. The origin of this turbulent motion and its impact on the cloud's lifetime
and star formation process is not well understood yet. It is however likely that large-scale disk
turbulence is the seed for turbulence in molecular clouds which again affects the star formation rate
that in turn drives again large scale disk turbulence and by this also the viscous secular evolution
of galactic disks.

\section{Summary}

We are currently living in a very exciting time where the various complex processes that can affect
galactic disk formation and evolution are being uncovered and studied observationally and theoretically.
Combined with the now well established cold dark matter structure formation scenario the time seems
ripe for self-consistent models of galaxy formation with predictive power. Given the
high capacities of present-day supercomputers it is understandable that one tries to
including as many processes as possible, most of which being however not
well understood. These models not only suffer from a large number of free parameters. They also
do not necessarily lead to insight as they are so complex and depend on so many different
implemented physical aspects that it is impossible to clearly understand what in the end the 
origin of a certain result will be.

I wonder whether one needs a high complexity in order to understand important questions 
of galactic disk evolution, like the two KS laws, the origin of turbulence in the diffuse 
interstellar medium or in molecular clouds or the origin of the strong correlation between the
viscous timescale and the star formation timescale.

Let us try to solve simple questions first before we focus on the complex puzzles that involve
many processes that are not well understood yet.

\end{document}